\documentclass[rmp,aps,twocolumn,nofootinbib]{revtex4}

\usepackage{graphicx,epic,eepic,epsfig,amsmath,latexsym,amssymb,verbatim,revsymb}

\usepackage{theorem}
\newtheorem{definition}{Definition}[section]

\newtheorem{lemma}[definition]{Lemma}

\newtheorem{theorem}[definition]{Theorem}
\newtheorem{corollary}[definition]{Corollary}

\def\squareforqed{\hbox{\rlap{$\sqcap$}$\sqcup$}}
\def\qed{\ifmmode\squareforqed\else{\unskip\nobreak\hfil
\penalty50\hskip1em\null\nobreak\hfil\squareforqed
\parfillskip=0pt\finalhyphendemerits=0\endgraf}\fi}
\def\endenv{\ifmmode\;\else{\unskip\nobreak\hfil
\penalty50\hskip1em\null\nobreak\hfil\;
\parfillskip=0pt\finalhyphendemerits=0\endgraf}\fi}
\newenvironment{proof}{\noindent \textbf{{Proof~} }}{\qed}

\newcommand{\outp}[2]{|#1\rangle\langle#2|}
\newcommand{\inp}[2]{\langle{#1}|{#2}\rangle} 
\mathchardef\ordinarycolon\mathcode`\:
\mathcode`\:=\string"8000
\def\vcentcolon{\mathrel{\mathop\ordinarycolon}}
\begingroup \catcode`\:=\active
  \lowercase{\endgroup
  \let :\vcentcolon
  }

\newcommand{\nc}{\newcommand}
\nc{\rnc}{\renewcommand}
\nc{\beq}{\begin{equation}}
\nc{\eeq}{{\end{equation}}}
\nc{\beqa}{\begin{eqnarray}}
\nc{\eeqa}{\end{eqnarray}}
\nc{\lbar}[1]{\overline{#1}}
\nc{\bra}[1]{\langle#1|}
\nc{\ket}[1]{|#1\rangle}
\nc{\ketbra}[2]{|#1\rangle\!\langle#2|}
\nc{\braket}[2]{\langle#1|#2\rangle}
\nc{\proj}[1]{| #1\rangle\!\langle #1 |}
\nc{\avg}[1]{\langle#1\rangle}
\rnc{\max}{\operatorname{max}}
\nc{\Rank}{\operatorname{Rank}}
\nc{\smfrac}[2]{\mbox{$\frac{#1}{#2}$}}
\nc{\tr}{\operatorname{Tr}}
\nc{\ox}{\otimes}
\nc{\dg}{\dagger}
\nc{\dn}{\downarrow}
\nc{\cA}{{\cal A}}
\nc{\cB}{{\cal B}}
\nc{\cC}{{\cal C}}
\nc{\cD}{{\cal D}}
\nc{\cE}{{\cal E}}
\nc{\cF}{{\cal F}}
\nc{\cG}{{\cal G}}
\nc{\cH}{{\cal H}}
\nc{\cI}{{\cal I}}
\nc{\cJ}{{\cal J}}
\nc{\cK}{{\cal K}}
\nc{\cL}{{\cal L}}
\nc{\cM}{{\cal M}}
\nc{\cN}{{\cal N}}
\nc{\cO}{{\cal O}}
\nc{\cP}{{\cal P}}
\nc{\cR}{{\cal R}}
\nc{\cS}{{\cal S}}
\nc{\cT}{{\cal T}}
\nc{\cX}{{\cal X}}
\nc{\cZ}{{\cal Z}}
\nc{\csupp}{{\operatorname{csupp}}}
\nc{\qsupp}{{\operatorname{qsupp}}}
\nc{\var}{\operatorname{var}}
\nc{\rar}{\rightarrow}
\nc{\lrar}{\longrightarrow}
\nc{\polylog}{\operatorname{polylog}}
\nc{\1}{{\openone}}

\newcommand{\setM}{\mathcal{M}}
\newcommand{\setX}{\mathcal{X}}

\nc{\RR}{{{\mathbb R}}}
\nc{\CC}{{{\mathbb C}}}
\nc{\FF}{{{\mathbb F}}}
\nc{\NN}{{{\mathbb N}}}
\nc{\ZZ}{{{\mathbb Z}}}
\nc{\PP}{{{\mathbb P}}}
\nc{\QQ}{{{\mathbb Q}}}
\nc{\UU}{{{\mathbb U}}}
\nc{\EE}{{{\mathbb E}}}
\nc{\id}{{\1}}
\newcommand{\bop}{\mathcal{B}}
\newcommand{\hil}{\mathcal{H}}
\newcommand{\states}{\mathcal{S}}
\newcommand{\assign}{:=}
\newcommand{\setS}{\mathcal{S}}
\newcommand{\setA}{\mathcal{A}}
\newcommand{\setB}{\mathcal{B}}

\nc{\be}{\begin{equation}}
\nc{\ee}{{\end{equation}}}
\nc{\bea}{\begin{eqnarray}}
\nc{\eea}{\end{eqnarray}}
\nc{\<}{\langle}
\rnc{\>}{\rangle}
\nc{\Hom}[2]{\mbox{Hom}(\CC^{#1},\CC^{#2})}
\nc{\rU}{\mbox{U}}

\nc{\ob}[1]{#1}

\nc{\snote}[1]{\textbf{(Steph:}#1\textbf{)}}
\nc{\mub}{b}
\nc{\mB}{\mathcal{B}}
\nc{\mS}{\mathcal{S}}
\nc{\Complex}{\mathbb{C}}
\nc{\Real}{\mathbb{R}}
\nc{\N}{\mbox{N}}
\nc{\MOLS}{\mbox{MOLS}}
\nc{\Natural}{\mathbb{N}}
\def\01{\{0,1\}}
\nc{\Tr}{\mbox{Tr}}

\begin{document}

\title{Entropic uncertainty relations -- A survey}

\author{Stephanie \surname{Wehner}}
\email[]{wehner@caltech.edu}
\affiliation{Institute for Quantum Information, Caltech, Pasadena, CA 91125, USA}
\author{Andreas \surname{Winter}}
\email[]{a.j.winter@bris.ac.uk}
\affiliation{Department of Mathematics, University of Bristol, Bristol BS8 1TW, U.K.}
\affiliation{Centre for Quantum Technologies, National University of Singapore,
 2 Science Drive 3, Singapore 117542}

\date{6 July 2009}

\begin{abstract}
Uncertainty relations play a central role in quantum mechanics. 
Entropic uncertainty relations in particular have gained significant 
importance within quantum information, providing the foundation for 
the security of many quantum cryptographic protocols. Yet, 
rather little is known about entropic uncertainty relations 
with more than two measurement settings. In this note we review 
known results and open questions.
\end{abstract}

\maketitle


The uncertainty principle is one of the fundamental ideas of quantum mechanics.
Since Heisenberg's uncertainty relations for canonically conjugate variables, 
they have been one of the most prominent examples of how quantum 
mechanics differs from the classical world~\cite{heisenberg:ur}.
Uncertainty relations today are probably best known
in the form given by~\cite{robinson:uncertainty}, who extended Heisenberg's result to two arbitrary observables
$A$ and $B$. Robertson's relation states that if we prepare many copies of the state $\ket{\psi}$, and measure
each copy individually using either $A$ or $B$, we have
\begin{align}\label{eq:heisenberg}
\Delta A \Delta B \geq \frac{1}{2} |\bra{\psi}[A,B]\ket{\psi}|
\end{align}
where $\Delta X = \sqrt{\bra{\psi}X^2\ket{\psi} - \bra{\psi}X\ket{\psi}^2}$ for $X \in \{A,B\}$ is
the standard deviation resulting from measuring $\ket{\psi}$ with observable $X$.
The consequence is the complementarity of quantum  mechanics: there is no way
to simultaneously specify definite values of non-commuting observables.
This, and later, formulations concern themselves with the tradeoff between the
``uncertainties'' in the value of non-commuting observables on the same state
preparation. In other words, they are comparing counterfactual situations.

It was eventually realized that other measures of ``spread'' of the distribution
on measurement outcomes can be used to capture the essence of uncertainty relations, which 
can be advantageous. 
Arguably the universal such measure is
the entropy of the distribution, which led Hirschmann to propose the first entropic uncertainty relation for position
and momentum observables~\cite{hirschmann:ur}. His results were later improved by the inequalities of~\cite{becker:ur}
and the uncertainty relations of~\cite{BB:M}, which we will review below.
In~\cite{BB:M} it is shown that this relation implies the Heisenberg uncertainty
relation~\eqref{eq:heisenberg}, and thus entropic uncertainty relations provide
us with a more general framework of quantifying ``uncertainty''.

That entropic uncertainty relations are indeed desirable was pointed out by~\cite{deutsch:ur}, 
who emphasized the fact that the lower bound
given by Robertson's uncertainty relation depends on the state $\ket{\psi}$.
In particular, this lower bound is trivial when $\ket{\psi}$ happens to give
zero expectation on $[A,B]$ -- which in finite dimension is always possible.
He addressed this problem by proving a first entropic uncertainty relation in terms
of the Shannon entropy for \emph{any} two non-degenerate observables, which gives a bound that
is \emph{independent} of the state to be measured. His uncertainty relation was later improved
by~\cite{maassen:ur}, following a conjecture by~\cite{kraus:ur}, which 
we will discuss in detail below. Apart from allowing to put universal lower
bounds on uncertainty even in finite dimension, another side effect of considering entropy
uncertainty relations is a conceptual liberation. Indeed, Robertson's inequality (\ref{eq:heisenberg})
is best when the right hand side is $\1$, i.e.~$A$ and $B$ are canonically conjugate
which happens if and only if they are related by a Fourier transform. In the
finite dimensional case, \cite{maassen:ur} show that the largest uncertainty
is obtained more generally for so-called \emph{mutually unbiased} observables,
which opens the way for uncertainty tradeoffs of more than two observables.
Even though entropic uncertainty relations thus play an 
important role in our understanding of quantum mechanics, 
and have interesting applications ranging from quantum cryptography~\cite{qkd:ur,serge:bounded}, 
information locking~\cite{barbara:locking}
to the question of separability~\cite{guehne:separable}, very little is known about them. 
Indeed, only in the case of two measurement settings do we have a reasonable 
understanding of such relations. 
The purpose of this review is to present what is known
about entropic uncertainty relations for a number of different entropic quantities. 

Let us first consider the general form of an entropic uncertainty relation more formally.
Let $\setM_j = \{M^x_j \mid M^x_j \in \bop(\hil)\}$ be a measurement on the space $\hil$ 
with a (finite) set of outcomes $x \in \setX$, that is, for all $x$ we have $M^x_j \geq 0$ and
$\sum_x M^x_j = \id$. For any quantum state $\rho$, the measurement $\setM_j$ induces a distribution $P_j$
over the outcomes given by $P_j(x) = \Tr(M^x_j\rho)$. We will write $H_\alpha(\setM_j|\rho)$ for an entropy $H_\alpha$ of
the resulting distribution. For example, for the Shannon entropy we have
$$
H(\setM_j|\rho) = - \sum_x \Tr(M^x_j \rho) \log \Tr(M^x_j \rho)\ .
$$
An entropic uncertainty relation captures the incompatibility of several measurements $\setM_1,\ldots,\setM_L$. In particular,
any such relation takes the form
\begin{align}\label{eq:genUR}
\mbox{for all } \rho \in \states(\hil) \qquad \frac{1}{L} \sum_{j=1}^L H_\alpha(\setM_j|\rho) \geq c_{\{\setM_j\}}\ ,
\end{align}
where $c_{\{\setM_j\}}$ is a constant depending solely on our choice of measurements, and not on the state $\rho$.
It is a particularly interesting question to find measurements for which $c_{\{\setM_j\}}$ is as large as possible.

\paragraph*{Outline.}
In Section~\ref{sec:prelim} we first provide an overview of the entropic quantities we will use throughout this text. 
We also introduce the concept of maximally strong uncertainty relations and discuss mutually unbiased bases, 
which play a special role in the study of uncertainty relations.
We then first consider the case of two measurement settings ($L=2$) in Section~\ref{sec:two} which is the only case well-understood. 
In Section~\ref{sec:multi} we then present an overview of the few results known for multiple measurements.
We conlude in Section~\ref{sec:apps} with some applications of uncertainty relations in cryptography.

\section{Preliminaries}\label{sec:prelim}

\subsection{Entropic quantities}
We begin by introducing all entropic quantities used in this text. The expert reader may safely skip this section.
Let $P_X$ be a distribution over a set $\setX$, where we write $P_X(x)$ for the probability of
choosing a particular element $x \in \setX$.
The \emph{R{\'e}nyi entropy}~\cite{renyi:entropy} of this distribution is defined as
$$
H_\alpha(P_X) = \frac{1}{1-\alpha} \log \left(\sum_{x \in \setX} P_X(x)^\alpha\right)\ ,
$$
for any $\alpha \geq 0$.
It will be useful to note that the R{\'e}nyi entropy is in fact related to the $\alpha$-norm of the vector $v$
of probabilities
$$
\|v\|_\alpha = \left(\sum_{x \in \setX} P_X(x)^{\alpha}\right)^{1/\alpha}
$$
by taking the logarithm
$$
H_\alpha(P_X) = \frac{\alpha}{1-\alpha} \log \|v\|_\alpha\ .
$$
A special case of the R{\'e}nyi entropy is the well-known Shannon entropy~\cite{shannon:entropy} obtained by 
taking the limit
$$
H(P_X) = \lim_{\alpha \rightarrow 1} H_\alpha(P_X) = - \sum_{x \in \setX} P_X(x) \log P_X(x)\ .
$$
We are especially interested in the so-called \emph{collision entropy}, that is, the R{\'e}nyi entropy of
order $\alpha =2$ given by
$$
H_2(P_X) = - \log \sum_{x \in \setX} P_X(x)^2\ ,
$$
and the \emph{min-entropy} given by the limit $\alpha \rightarrow \infty$ as
$$
H_\infty(P_X) = - \log \max_{x \in \setX} P_X(x)\ .
$$
The R{\'e}nyi entropies are monotonically decreasing in $\alpha$, i.e.
$$
H_\alpha(\cdot) \geq H_\beta(\cdot)\ ,
$$
for $\alpha \leq \beta$. In particular, we thus have $H_\infty(\cdot) \leq H_2(\cdot) \leq H(\cdot)$.
Note that any such entropies can take on values in the interval $0 \leq H_\alpha(\cdot) \leq \log |\setX|$,
where the lower bound is clearly attained if the distribution is sharply defined with $P_X(x) = 1$ for some $x \in \setX$, 
and the upper bound is attained when $P_X(x) = 1/|\setX|$ is the uniform distribution.

In the following, we will write 
$$
H_\alpha(\mB|\rho) \assign H_\alpha(\{\proj{x}\}|\rho)
$$ 
to denote the entropy
arising from a measurement in an orthonormal basis $\mB = \{\ket{x}\mid x \in [d]\}$ and use
$$
H_\alpha(\setA|\rho) \assign H_\alpha(\{A_x\}|\rho)
$$ 
to denote the entropy arising from measuring with observables $\setA$ given by the projectors $\{A_x\}$.

\subsection{Maximally strong uncertainty relations}

An intriguing question is to find measurements which are very incompatible, in the sense that the r.h.s of~\eqref{eq:genUR}
is very large. We will refer to this as a \emph{strong uncertainty relation}.
Note that given any set of projective measurements $\setM_1,\ldots,\setM_L$, we can always find a state $\rho$
such that
$$
H_\alpha(\setM_j|\rho) = 0
$$
for one of the measurements $\setM_j$, namely by choosing $\rho$ to be an eigenstate of one of the measurement operators.
We thus know that the r.h.s of~\eqref{eq:genUR} can never exceed
$$
\log |\setX| (1 - 1/L) \geq c_{\{\setM_j\}} \geq 0\ .
$$
If for any choice of measurements the lower bound is given by $c_{\{\setM_j\}} = \log |\setX| (1-1/L)$, we know that if $\rho$ has zero entropy for
one of the measurements, the entropy is maximal for all others. We call a set of measurements that satisfy this
property \emph{maximally incompatible}, and refer to the corresponding uncertainty relation as being \emph{maximally strong}.
As outlined below, mutually unbiased bases lead to maximally strong uncertainty relations for $L=2$ measurements.
This however does not hold in general for the case of $L > 2$. We will also see that maximally incompatible measurements
can be found for any $L$ if we only consider $|\setX| = 2$ outcomes.

For measurements in different~\emph{bases}, note that all bases must be mutually unbiased in order for us to obtain strong uncertainty relations:
Suppose two bases $\mB_1$ and $\mB_2$ are not mutually unbiased, and there exist two basis vectors $\ket{x} \in \mB_1$ and $\ket{y} \in \mB_2$
that have higher overlap $|\inp{x}{y}|^2 > 1/d$. Then choosing $\rho = \proj{x}$ yields zero entropy when measured in basis $\mB_1$ and less
than full entropy when measured in the basis $\mB_2$.

\subsection{Mutually unbiased bases}\label{sec:mub}

Since mutually unbiased bases play an important role in the study of uncertainty relations, we briefly review two well-known constructions
for which particular uncertainty relations are known to hold.
\begin{definition}[MUBs] \label{def-mub}
Let $\mathcal{B}_1 = \{\ket{\mub^1_1},\ldots,\ket{\mub^1_{d}}\}$ and $\mB_2 =
\{\ket{\mub^2_1},\ldots,\ket{\mub^2_{d}}\}$ be two orthonormal bases in
$\Complex^d$. They are said to be
\emph{mutually unbiased}  if
$|\inp{\mub^1_k}{\mub^2_l}| = 1/\sqrt{d}$, for every $k,l \in[d]$. A set $\{\mathcal{B}_1,\ldots,\mathcal{B}_m\}$ of
orthonormal bases in $\Complex^d$ is called a \emph{set of mutually
unbiased bases} if each pair of bases is mutually unbiased.
\end{definition}

For example, the well-known computational and Hadamard basis are mutually unbiased.
We use $N(d)$ to denote the maximal number of MUBs in dimension $d$.
In any dimension $d$, we have that
$\N(d) \leq d+1$~\cite{boykin:mub}. If $d = p^k$ is a prime power, we have
that $\N(d) = d+1$ and explicit constructions are
known~\cite{boykin:mub,wootters:mub}. If $d = s^2$ is a square,
$\N(d) \geq \MOLS(s)$ where $\MOLS(s)$ denotes the number of mutually orthogonal
$s \times s$ Latin squares~\cite{wocjan:mub}.
In general, we have
$\N(n m) \geq \min\{\N(n),\N(m)\}$ for all $n,m \in \Natural$~\cite{zauner:diss,klappenecker:mubs}.
From this it follows that in any dimension, 
there is an explicit construction for 3 MUBs~\cite{grassl:mub}.
Unfortunately, not much else is known. For example,
it is still an open problem whether there exists a set of $7$ 
(or even $4$!) MUBs in dimension $d=6$.
In this text, we refer to two specific constructions of mutually unbiased bases.
There exists a third construction based on Galois rings~\cite{klappenecker:mubs}, which we do not
consider here, since we do not know of any specific uncertainty relations in this setting.

\subsubsection{Latin squares}
First, we consider MUBs based on mutually orthogonal Latin squares~\cite{wocjan:mub}.
Informally, an $s \times s$ Latin square over the symbol set $[s]$
is an arrangement
of elements of $[s]$ into an $s \times s$ square such that in each row and each column every element
occurs exactly once. Let $L_{ij}$ denote the entry in a Latin square in row $i$ and column $j$.
Two Latin squares $L$ and $L'$ are called mutually orthogonal if and only if
$\{(L_{i,j},L'_{i,j})|i,j \in [s]\} = \{(u,v)|u,v \in [s]\}$. Intuitively, this means that if we place
one square on top of the other, and look at all pairs generated by the overlaying elements, all possible pairs
occur. An example is given in Figure~\ref{LatinSquareMOLS} below.
From any $s\times s$ Latin square we can obtain a basis for $\Complex^{s}\otimes \Complex^{s}$.
First, we construct $s$ of the basis vectors from the entries of
the Latin square itself. Let
$$
\ket{v_{1,\ell}} = \frac{1}{\sqrt{s}} \sum_{i,j\in [s]} E^L_{i,j}(\ell) \ket{i,j},
$$
where $E^L$ is a predicate such that $E^L_{i,j}(\ell) = 1$ if and only if $L_{i,j} = \ell$.
Note that for each $\ell$ we have exactly $s$ pairs $i,j$ such that $E_{i,j}(\ell) = 1$, because
each element of $[s]$ occurs exactly $s$ times in the Latin square.
Secondly, from each such vector we obtain $s-1$ additional vectors by adding successive rows
of an $s \times s$ complex Hadamard matrix $H = (h_{ij})$ as coefficients to obtain the remaining
$\ket{v_{t,j}}$ for $t \in [s]$, where $h_{ij} = \omega^{ij}$ with $i,j \in \{0,\ldots,s-1\}$ and
$\omega = e^{2 \pi i/s}$.
Two additional MUBs can then be obtained in the same way from the two non-Latin squares where
each element occurs for an entire row or column respectively. From each mutually orthogonal Latin square
and these two extra squares which also satisfy the above orthogonality condition, we obtain one basis.
This construction therefore gives $\MOLS(s) + 2$ many MUBs. It is known that if $s = p^k$ is a
prime power itself, we obtain
$p^k+1\approx \sqrt{d}$ MUBs from this construction. Note, however, that there do exist many more
MUBs in prime power dimensions, namely $d+1$. If $s$ is not a prime power, it is merely known
that $\MOLS(s) \geq s^{1/14.8}$~\cite{wocjan:mub}.

\begin{figure}[h]
\begin{minipage}{0.45\textwidth}
\begin{center}
\includegraphics[width=2cm]{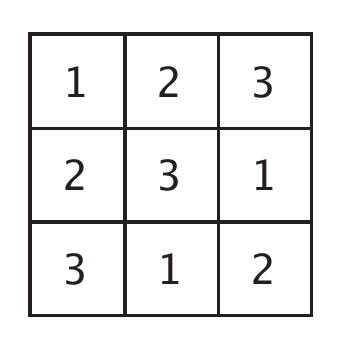}
\end{center}
\end{minipage}
\begin{minipage}{0.45\textwidth}
\begin{center}
\includegraphics[width=2cm]{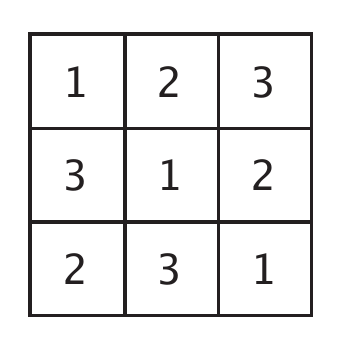}
\end{center}
\end{minipage}
\caption{Mutually orthogonal latin squares}
\label{LatinSquareMOLS}
\end{figure}

As an example, consider the first $3 \times 3$ Latin square depicted in Figure~\ref{LatinSquareMOLS} and
the $3 \times 3$ complex Hadamard matrix\index{complex Hadamard matrix}
$$
H = \left(\begin{array}{ccc}
    1 &1& 1\\
    1 &\omega &\omega^2\\
    1 &\omega^2& \omega
\end{array}\right),
$$
where $\omega = e^{2 \pi i/3}$.
First, we obtain vectors
\begin{eqnarray*}
\ket{v_{1,1}} &=& \frac{1}{\sqrt{3}}(\ket{1,1} + \ket{2,3} + \ket{3,2})\ ,\\
\ket{v_{1,2}} &=& \frac{1}{\sqrt{3}}(\ket{1,2} + \ket{2,1} + \ket{3,3})\ ,\\
\ket{v_{1,3}} &=& \frac{1}{\sqrt{3}}(\ket{1,3} + \ket{2,2} + \ket{3,1})\ .
\end{eqnarray*}
With the help of $H$ we obtain 3
additional vectors from the ones above. From the vector $\ket{v_{1,1}}$,
for example, we obtain
\begin{eqnarray*}
\ket{v_{1,1}} &=& \frac{1}{\sqrt{3}}(\ket{1,1} + \ket{2,3} + \ket{3,2})\ ,\\
\ket{v_{2,1}} &=& \frac{1}{\sqrt{3}}(\ket{1,1} + \omega \ket{2,3} + \omega^2 \ket{3,2})\ ,\\
\ket{v_{3,1}} &=& \frac{1}{\sqrt{3}}(\ket{1,1} + \omega^2 \ket{2,3} + \omega \ket{3,2})\ .
\end{eqnarray*}
This gives us basis $\mathcal{B} = \{\ket{v_{t,\ell}}|t,\ell \in [s]\}$ for $s = 3$.
The construction of another basis follows in exactly the same way from a mutually orthogonal
Latin square. The fact that two such squares $L$ and $L'$ are mutually orthogonal ensures
that the resulting bases will be mutually unbiased. Indeed, suppose we are given another such basis,
$\mathcal{B'} = \{\ket{u_{t,\ell}}|t,\ell \in [s]\}$ belonging to $L'$. We then have for any $\ell,\ell' \in [s]$ that
$|\inp{u_{1,\ell'}}{v_{1,\ell}}|^2 =
|(1/s) \sum_{i,j\in [s]} E^{L'}_{i,j}(\ell') E^L_{i,j}(\ell)|^2 = 1/s^2$, as there exists exactly only
one pair $\ell,\ell' \in [s]$ such that $E^{L'}_{i,j}(\ell') E^L_{i,j}(\ell) = 1$. Clearly, the same argument
holds for the additional vectors derived from the complex Hadamard matrix.

\subsubsection{Generalized Pauli matrices}

The second construction we consider is based on the generalized Pauli matrices
$X_d$ and $Z_d$~\cite{boykin:mub}, defined by their actions on the
computational basis $C = \{\ket{0},\ldots,\ket{d-1}\}$ as follows:
\begin{eqnarray*}
X_d\ket{k} &=& \ket{k+1\mod d}\\
Z_d\ket{k} &=& \omega^k\ket{k},~\forall \ket{k} \in C\ ,
\end{eqnarray*}
where $\omega = e^{2 \pi i/d}$.  We say that $\left(X_{d}\right)^{a_1} \left(Z_{d}\right)^{b_1} 
\otimes \cdots \otimes \left(X_{d}\right)^{a_N} \left(Z_{d}\right)^{b_N}$ for
$a_k,b_k \in \{0,\ldots,d-1\}$ and $k \in [N]$ is a \emph{string of Pauli matrices}.
Note that for $d=2$ these are just the usual Pauli matrices.

If $d$ is a prime, it is known that the $d+1$ MUBs constructed first by
Wootters and Fields~\cite{wootters:mub} can also be obtained as the eigenvectors of the
matrices $Z_d,X_d,X_dZ_d,X_dZ_d^2,\ldots,X_dZ_d^{d-1}$~\cite{boykin:mub}. If $d = p^k$ is a prime power,
consider all $d^2-1$ possible strings of Pauli matrices excluding the identity and group them
into sets $C_1,\ldots,C_{d+1}$ such that $|C_i| = d - 1$ and $C_i \cap C_j = \{\id\}$ for $i \neq j$ and all
elements of $C_i$ commute. Let $B_i$ be the common eigenbasis of all elements of $C_i$. Then
$B_1,\ldots,B_{d+1}$ are MUBs~\cite{boykin:mub}. A similar result for $d = 2^k$ has also been shown
in~\cite{lawrence:mub}.
A special case of this construction are the three mutually unbiased bases in dimension $d=2^k$ given by
the unitaries $\id^{\otimes k}$, $H^{\otimes k}$ and $K^{\otimes k}$
applied to the computational basis, where $H$ is the Hadamard transform and $K = (\id + i \sigma_x)/\sqrt{2}$.
A simple example of this construction are the mutually unbiased bases in dimension $d=2$ which are
given by the eigenvectors of the Pauli matrices $X$, $Z$ and $Y$.
A very interesting aspect of such mutually unbiased bases is that there exists an ordering $\mB_1,\ldots,\mB_{d+1}$ 
and a unitary $U$ that cyclically permutes all bases, that is, $U \mB_{j} = U\mB_{j+1}$ for all $j$, where 
$U \mB_{d+1} = \mB_1$~\cite{wootters:mur}.

\section{Two measurements}\label{sec:two}

The case of two measurements ($L=2$) is reasonably well understood in any dimension, and for any number of outcomes.
This case was of particular interest as is directly inspired by the two measurements for which Heisenberg
had originally formulated his uncertainty relation, i.e., position and momentum.
We begin by recalling some of the history of this fascinating problem, before reviewing the currently relevant results.

\subsection{History}

The first entropic uncertainty relation was given by~\cite{hirschmann:ur} for position and momentum observables, which
was improved by the inequalities of~\cite{becker:ur} and the entropic uncertainty relations of~\cite{BB:M} to
an entropic uncertainty relation for systems of $n$ canonical pairs
of position and momentum coordinates $X_i$ and $P_i$:
\[
  H(X_1\ldots X_n | \rho) + H(P_1\ldots P_n | \rho) \geq n \log(e\pi)\ ,
\]
where $H(Q_1\ldots Q_n | \varphi)$ refers to the (differential)
Shannon entropy of the joint distribution of the coordinates
$Q_1,\ldots Q_n$ when measured on the state $\rho$.

That entropic uncertainty relations are of great importance was pointed out by~\cite{deutsch:ur}, who 
proved that for measurements in two bases $\setA$ and $\setB$ we have
$$
\frac{1}{2}(H(\setA|\rho) + H(\setB|\rho)) \geq - \log\left(\frac{1 + c(\setA,\setB)}{2}\right)\ ,
$$
where $c(\setA,\setB) \assign \max\{|\inp{a}{b}| \mid \ket{a} \in \setA, \ket{b} \in \setB\}$.
We will see later that the same bound holds for the min-entropies $H_\infty(\cdot)$. 
His results were extended to a continuous setting for angle-angular momentum and position and momentum
by~\cite{partovi:angleUR}, which in turn was improved by~\cite{birula:angleUR}. Different relations
for particular angular momentum observables were later also derived by~\cite{birula:angularNew}.
A R{\'e}nyi entropic version of such an uncertainty relation may be found in~\cite{birula:new}.


\subsection{Measurements in different bases}

\subsubsection{Any choice of bases}
Following a conjecture by Kraus~\cite{kraus:ur}, Maassen and Uffink~\cite{maassen:ur} improved Deutsch's uncertainty relation for
measurements in two different bases. In particular, they showed that if we measure any state $\rho \in \hil$ with $\dim \hil = d$
using observables with orthonormal eigenbases $\setA = \{\ket{a_1},\ldots,\ket{a_d}\}$ and
$\setB = \{\ket{b_1},\ldots,\ket{b_d}\}$ respectively, we have
\begin{align}\label{eq:muBound}
\frac{1}{2}\bigl( H(\setA|\ket{\psi}) + H(\setB|\ket{\psi}) \bigr) \geq - \log c(\setA,\setB)\ ,
\end{align}
where $c(\setA,\setB) \assign \max\{|\inp{a}{b}| \mid \ket{a} \in \setA, \ket{b} \in \setB\}$.
Since $H(\cdot)$ is concave in $\ket{\psi}$, this result also applies to mixed states $\rho$.
What is the strongest possible relation we could obtain? That is, which choices of $\setA$ and $\setB$ maximize 
the r.h.s. of equation (\ref{eq:muBound})?
It turns out that the maximum is reached when the two bases
are mutually unbiased (see Section~\ref{sec:mub})
i.e.~when all
the inner products on the right hand side above are equal to $1/\sqrt{d}$. We then
obtain
that the entropy sum is lower bounded by $\frac{1}{2}\log d$. This is tight, as the
example of $\ket{\varphi} = \ket{a_1}$ shows.
Note that for general observables, this lower bound is
not necessarily tight, but its usefulness lies in the fact that it is in terms of \emph{very simple} geometric
information of the relative position of the bases. 

\subsubsection{Improved bounds for specific bases}
For dimension $d=2$ optimal uncertainty relations have been obtained
for two observables $A = \vec{a} \cdot \vec{\sigma}$ and $B = \vec{b} \cdot \vec{\sigma}$ where $\vec{\sigma} = (X,Y,Z)$,
for some angles of the Bloch vectors $\vec{a}\cdot\vec{b}$ analytically, and for others numerically~\cite{ghirardi:ur}.
Uncertainty relations which give improved bounds for a large class of measurements in two different bases $\setA$ and $\setB$ have also 
been obtained in~\cite{sanchez:evenBetter} for the case that the overlap between two basis vectors is large, that is, $c(\setA,\setB) \geq 1/\sqrt{2}$.
Letting $c \assign c(\setA,\setB)$, the following analytical bound is shown for this regime
\begin{align*}
\frac{1}{2}\bigl( H(\setA|\rho) &+ H(\setB|\rho) \bigr)\\
 &\geq - \frac{1+c}{2} \log\left(\frac{1+c}{2}\right) - \frac{1-c}{2} \log\left(\frac{1-c}{2}\right)\ ,
\end{align*}
and a numerical bound is provided that is slightly better for $1/\sqrt{2} \leq c \leq 0.834$. 

\subsubsection{Relations for R{\'e}nyi entropies}
It is an often overlooked fact that Maassen and Uffink actually also show uncertainty relations in terms
of the R{\'e}nyi entropies. In particular, they extend a result by~\cite{landau:ur} to show that
for any $\ket{\psi}$
$$
\frac{1}{2} \bigl(H_\infty(\setA|\ket{\psi}) + H_\infty(\setB|\ket{\psi}) \bigr) \geq - \log \left[\frac{1 + c(\setA,\setB)}{2}\right]\ .
$$
To see that this bound can be tight for some choices of $\setA$ and $\setB$, consider two mutually unbiased bases
in dimension $d=2$. For example, the computational $\setA = \{\ket{0},\ket{1}\}$ and the Hadamard basis 
$\setB = \{\ket{+},\ket{-}\}$. The lower bound
then becomes $- \log(1/2 + 1/(2\sqrt{2}))$, which is attained for $\ket{\psi} = \cos(\pi/8) \ket{0} + 
\sin(\pi/8)\ket{1}$.
They furthermore use a result for $\alpha$-norms~\cite{riesz:norms} to show that
the following relation holds in terms for R{\'e}nyi entropies of order $\alpha$ and $\beta$
satisfying $\alpha > 1$ and $\beta = \alpha/(2\alpha-1) < 1$
$$
\frac{1}{2} \bigl(H_\alpha(\setA|\ket{\psi}) + H_\beta(\setB|\ket{\psi})\bigl) \geq - \log c(\setA,\setB)\ ,
$$
for any state $\ket{\psi}$, which gave the result for the Shannon entropy above 
in the limit of $\alpha, \beta \rightarrow 1$.

\subsection{General measurements}

\subsubsection{Shannon entropy}
The result by~\cite{maassen:ur} has been extended to the case of a general POVM.
The first such result was given by~\cite{hall:ur}, who pointed out that their result can easily be extended to the case of rank one 
POVMs. His result was subsequently strengthened~\cite{massar:ur,rastegin:errMassar} by noting that any two 
POVMS $\setM_1 = \{\proj{x_1}\mid \ket{x_1} \in \hil\}$ and $\setM_2 = \{\proj{x_2} \mid \ket{x_2} \in \hil\}$ acting on the Hilbert
space $\hil$ have a Naimark extension to an ancillary space $\hil_{\rm anc}$ such that $U \ket{\tilde{x}_1} = \ket{x_1} + U \ket{\tilde{x}_1}$, 
and $U\ket{\tilde{x}_2} = \ket{x_2} + U \ket{\hat{x}_2}$ 
for any unitary $U = (\id_\hil \oplus V_{\rm anc})$ acting only on $\hil_{\rm anc}$,
where $\{\ket{\hat{x}_1}, \ket{\hat{x}_2} \in \hil_{\rm anc}\}$ form an orthonormal bases 
on the ancillary system.
Maximizing over such unitaries, that is, possible extensions of the POVM, one obtains the bound
\begin{align*}
\frac{1}{2} (H(\setM_1|\rho) &+ H(\setM_2|\rho))\\
&\geq \max_{U} - \log \max_{x,y} |\bra{\tilde{x}_1}U\ket{\tilde{y}_2}| 
\end{align*}
any state $\ket{\psi} \in \hil$. The general setting was analyzed by~\cite{krishna:ur} who showed that
\begin{align*}
\frac{1}{2} (H(\setM_1|\rho) &+ H(\setM_2|\rho))\\
&\geq - \log \max_{x,y}||(M_1^{(x)})^{1/2} (M_2^{(y)})^{1/2}||
\end{align*}
for any POVMS $\setM_1 = \{M_1^{(x)}\mid M_1^{(x)} \in \bop(\hil)\}$ and $\setM_2 = \{M_2^{(y)}\mid M_2^{(y)} \in \bop(\hil)\}$ and any 
state $\ket{\psi} \in \hil$.

\subsubsection{R{\'e}nyi entropy}
Entropic uncertainty relations for R{\'e}nyi entropies have also been obtained for the case of POVMs.
In particular, it has been shown by~\cite{rastegin:ur1,rastegin:ur2} that
for any two POVMS $\setM_1$ and $\setM_2$ and any state
$$
\rho = \sum_j \lambda_j \proj{\psi_j}\ ,
$$
we have
\begin{align*}
\frac{1}{2} &(H_\alpha(\setM_1|\rho) + H_\beta(\setM_2|\rho))\\
& \geq - \log \left[ \max_{j,x,y} \frac{|\bra{\psi_j}M_1^{(x)} M_2^{(y)}\ket{\psi_j}|}
{||(M_1^{(x)})^{1/2} \ket{\psi_j}|| ||(M_2^{(y)})^{1/2}\ket{\psi_j}||} \right]\ ,
\end{align*}
for $1/\alpha + 1/\beta = 2$.


\subsection{Beyond classical entropies}
\label{subsec:information}
In the context of quantum information theoretical applications
some other uncertainty relations were discovered, which are entropic in spirit,
but lie outside of the formalism introduced above. 

Here we quote two, which can be viewed as extensions of the
inequality of~\cite{maassen:ur} in the case of two measurement bases 
related by the Fourier transform, to multipartite quantum systems
and involving the von Neumann entropy $S(\rho) = -\tr\rho\log\rho$.
With this entropy, one can formally construct a mutual information
and a conditional entropy, respectively, for bipartite states $\rho_{AB}$
with marginals $\rho_A = \tr_B \rho_{AB}$ and $\rho_B = \tr_A \rho_{AB}$:
\begin{align*}
  I(A:B) &= I(A:B)_{\rho} := S(\rho_A) + S(\rho_B) - S(\rho_{AB}), \\
  S(A|B) &= S(A|B)_\rho   := S(\rho_{AB}) - S(\rho_B),
\end{align*}
Both inequalities compare two conjugate bases, i.e.~without loss of
generality, one is the standard basis $\left\{ \ket{z}:z=0,\ldots,d-1 \right\}$,
the other one its Fourier transform 
$\left\{ \ket{\hat{x}} = \sum_z e^{2\pi i xz/d}\ket{z}:x=0,\ldots d-1 \right\}$.
(These are just the eigenbases of the generalized $Z$ and $X$ Pauli operators.) 
Denote the projections onto these bases by $\cZ$, $\cX$, respectively:
\begin{align*}
  \cZ(\rho) &= \sum_z \proj{z} \rho \proj{z}, \\
  \cX(\rho) &= \sum_x \proj{\hat{x}} \rho \proj{\hat{x}}.
\end{align*}

The first uncertainty relation is by~\cite{CW05}: For a bipartite quantum
state $\rho_{AB}$ such that $\rho_A$ is maximally mixed,
\begin{equation}
  I(A:B)_{\cZ\ox\textrm{id}(\rho)} + I(A:B)_{\cX\ox\textrm{id}(\rho)} \leq I(A:B)_\rho.
\end{equation}

The second is by~\cite{RenesBoileau:inequality}, who show similarly
that for any tripartite state $\rho_{ABC}$,
\begin{equation}
  S(A|B)_{\cZ\ox\textrm{id}(\rho)} + S(A|C)_{\cX\ox\textrm{id}(\rho)} \geq \log d.
\end{equation}
Note that this directly reduces to (\ref{eq:muBound}) for trivial systems
$B$ and $C$ -- which is why \cite{RenesBoileau:inequality} conjecture the
following inequality when $\cZ$ and $\cX$ are more generally the projections onto
two arbitrary bases $\cA$ and $\cB$, respectively:
\[
  S(A|B)_{\cZ\ox\textrm{id}(\rho)} + S(A|C)_{\cX\ox\textrm{id}(\rho)} \geq - \log c(\cA,\cB).
\]

\section{More than two measurements}\label{sec:multi}

We now review the known results for entropic uncertainty relations for more than two measurement settings. 
Rather little is known in this scenario, except for a number of special cases. In particular, 
it is an interesting open question whether strong uncertainty relations even exist for a small constant number of measurement settings and more than two measurement outcomes.
As pointed out already in the beginning, this is conceivable because unlike canonically
conjugate variables, which come in pairs, there are generally more than two 
mutually unbiased observables.

\subsection{Random choice of bases}

First of all, it may not be at all obvious that strong uncertainty relations can even be obtained at all for more than two measurement settings, 
independent of the number of measurement outcomes. 
We will use $\mB_j = \{U_j \ket{x} \mid x \in \{0,\ldots,d-1\}\}$ where $\ket{x}$ forms an orthonormal basis
for $\hil$ to denote the basis obtained by rotating the standard basis into the basis determined by the unitary $U_j$.
It was shown in~\cite{rand} that $L=(\log d)^4$
unitaries $U_j$ 
chosen from the Haar measure
randomly and independently 
obey
$$
\frac{1}{L} \sum_{j=1}^L H(\mB_j|\rho) \geq \log d - O(1) 
                                          = (\log d)\left( 1-O\left(\frac{1}{\log d}\right) \right)
$$ 
with high probability, and for sufficiently large dimension $d$. 
It is important to note that the number of measurement settings is not constant but depends on the dimension. 

\subsection{Mutually unbiased bases}

Now that we know that it is in principle possible to obtain reasonably 
strong uncertainty relations, can we construct explicit measurements 
for which we obtain such relations? Recall that it is a necessary 
condition for bases to be mutually unbiased in order to obtain a maximally
strong uncertainty relation in the first place. 
Given the fact that if we have two measurement settings, choosing the measurement bases to be mutually unbiased leads 
to maximally strong uncertainty relations, it may be tempting to conclude that choosing our measurements to be mutually unbiased is in general also a
sufficient condition.
Perhaps surprisingly, this is not the case.

\subsubsection{For $d+1$ mutually unbiased bases}

We first consider the case of all $d+1$ mutually unbiased bases, for which we \emph{can} obtain strong uncertainty relations.
In particular, \cite{sanchez:old, ivanovic:ur}
has shown that for the mutually unbiased bases $\mB_1,\ldots,\mB_{d+1}$ we have for any state $\rho$
\begin{align}\label{eq:mubAll}
\frac{1}{d+1} \sum_{j=1}^{d+1} H(\mB_j|\rho) \geq 
\log(d+1)-1\ .
\end{align}
If the dimension $d$ is even, this can further be improved to~\cite{sanchez:improved}
\begin{align*}
\frac{1}{d+1} \sum_{j=1}^{d+1} H(\mB_j|\rho) 
                    &\geq \frac{1}{d+1} \left[\frac{d}{2} \log\left(\frac{d}{2}\right)\right.\\
                    &\phantom{=====:}
                     + \left.\left(\frac{d}{2} + 1\right) \log\left(\frac{d}{2} + 1\right)\right]\ .
\end{align*}
In dimension $d=2$, the latter bound gives $2/3$, which is tight for the mutually unbiased bases given by the 
eigenvectors of the Pauli matrices $X$, $Z$ and $Y$. 
The case of $d=2$ was also addressed separately in~\cite{sanchez:entropyD2}.

It is worth noting that the first bound~\eqref{eq:mubAll} is in fact obtained by 
first lower bounding the Shannon entropy $H(\cdot)$ 
by the collision entropy $H_2(\cdot)$, and then one proves that
\begin{align}\label{eq:mubAllH2}
\frac{1}{d+1} \sum_{j=1}^{d+1} H_2(\mB_j|\rho) \geq 
\log(d+1)-1\ .
\end{align}
This inequality can also be proven using the fact that a full set of mutually unbiased bases forms a $2$-design~\cite{BallesterWehner},
and we provide a completely elementary proof of this inequality in the appendix. Interestingly, it has been shown~\cite{wootters:mur} that
the states $\rho$ minimizing the l.h.s of~\eqref{eq:mubAllH2} are states which are invariant under a unitary transformation 
that permutes the mutually unbiased bases as discussed in Section~\ref{sec:mub}.

\subsubsection{For less than $d+1$ mutually unbiased bases}
What about less than $d+1$ mutually unbiased bases?
First of all, note that it is easy to see that we do not always obtain a maximally strong uncertainty relation
in this setting. Consider dimension $d=3$ and three mutually unbiased bases $\mB_1$, $\mB_2$ and $\mB_3$ given by the eigenvectors
of $X_3$, $Z_3$ and $X_3 Z_3$ respectively. Then a simple calculation shows that for example for the state
$\ket{\psi} = (\ket{1} - \ket{2})/\sqrt{2}$ we have $H(\mB_j|\ket{\psi}) = 1$ for all bases $j \in \{1,2,3\}$ and hence
$$
\frac{1}{3}\sum_{j=1}^3 H(\mB_j|\ket{\psi}) = 1 < \frac{2}{3} \log 3\ .
$$
In~\cite{barbara:locking} (see the eprint version) numerical work on three and more
mutually unbiased bases in prime dimensions up to $29$ is reported, which
are consistent with a behavior of $1-O(1/k)$ of $h(k)$. The mutually
unbiased bases are taken as a subset of the MUBs constructed via the
generalized Pauli matrices in prime power dimension.

Trivial bounds for more than two and less than $d+1$ can be derived quite easily. For example, for any number 
of mutually unbiased bases $\mB_1, \ldots,\mB_L$ we obtain 
by combining~\eqref{eq:muBound} for each pair of bases $\mB_i$ and $\mB_j$ that
\begin{align}\label{eq:mubTrivial}
\frac{1}{L} \sum_{j=1}^L H(\mB_j|\rho) \geq \frac{\log d}{2}\ .
\end{align}
As shown in the appendix, it is also easy to see that
$$
\frac{1}{L}\sum_{j=1}^L H(\mB_j|\rho) \geq - \log \frac{L + d - 1}{d L}\ .
$$

Curiously, it turns out~\cite{BallesterWehner} that in square prime power dimensions $d = p^{2 \ell}$
there
exist up to $L=p^\ell+1$ MUBs derived from the generalized Pauli matrices 
for which we obtain extremely weak uncertainty relations! In particular, we have
for any such set of MUBs that the lower bound of~\eqref{eq:mubTrivial} can be attained\footnote{And many more if one relaxes
the condition of mutual unbiasedness to approximate unbiasedness, using
the techniques of~\cite{rand}.}, that is, 
$$
\min_\rho \frac{1}{L} \sum_j H(\mB_j|\rho) = \frac{\log d}{2}\ .
$$
Furthermore, the same is true for \emph{all} mutually unbiased bases derived from Latin squares.
These results clearly show that mutual unbiasedness is \emph{not} enough to obtain strong uncertainty relations.
Combined with the numerical results from above, we also note that the dimension $d$, as well as the choice of mutually unbiased bases may  
indeed matter. In~\cite{BallesterWehner} it was noted that the set of mutually unbiased bases derived from the generalized Pauli matrices
for which we obtain weak uncertainty relations
are exactly those which are separable across the space $\Complex^{p^\ell} \otimes \Complex^{p^\ell}$. However, we can now conclude from
the results of~\cite{wootters:mur} that there is nothing inherently special about these separable bases, since there exists a unitary $U$
that maps them to a set of entangled bases (see Section~\ref{sec:mub}).
It has also been shown by~\cite{andris} that for any three
bases from the ``standard'' mutually unbiased bases construction in
prime power dimension
the lower bound cannot exceed $\left( \frac{1}{2}+o(1) \right) \log d$,
for large dimension, and assuming the Generalized Riemann Hypothesis.
Furthermore, for any $0\leq \epsilon\leq 1/2$,
there always exist $k=d^\epsilon$ many of these bases such
that the lower bound cannot be larger than $\left( \frac{1}{2}+\epsilon+o(1) \right)\log d$.
It remains an interesting open question to show tight uncertainty relations for all mutually unbiased bases.

\subsection{Anti-commuting observables}

Maximally strong uncertainty relations are known to exist for any number of measurement settings $L$, if we
limit ourselves to $\log |\setX| = 2$ outcomes.
These uncertainty relations are derived for generators of a Clifford algebra~\cite{lounesto:book,dietz:blochsphere},
which has many beautiful geometrical properties.
For any integer $n$, the free real associative algebra generated by
$\Gamma_1,\ldots,\Gamma_{2n}$, subject to the anti-commutation
relations
\begin{equation}
  \label{eq:anti}
  \{ \Gamma_j,\Gamma_k \} = \Gamma_j\Gamma_k + \Gamma_k\Gamma_j = 2\delta_{jk} \1,
\end{equation}
is called \emph{Clifford algebra}. It has a unique representation by
Hermitian matrices on $n$ qubits (up to unitary equivalence).
This representation can be obtained via the famous 
Jordan-Wigner transformation~\cite{JordanWigner}:
\begin{align*}
  \Gamma_{2j-1} &= Z^{\ox(j-1)} \ox X \ox \1^{\ox(n-j)}, \\
  \Gamma_{2j}   &= Z^{\ox(j-1)} \ox Y \ox \1^{\ox(n-j)},
\end{align*}
for $j=1,\ldots,n$, where we use $X$, $Y$ and $Z$ to denote the Pauli matrices.
An additional such matrix can be found by taking the product
$\Gamma_0 := \Gamma_1\Gamma_2\ldots\Gamma_{2n}$, which is sometimes known as the pseudo-scalar.
To see how such operators are observables with two measurement outcomes, note that the eigenvalues of $\Gamma_i$ always come in pairs:
Let $\ket{\eta}$ be an eigenvector of $\Gamma_i$ with eigenvalue $\lambda$.
From $\Gamma_i^2 = \id$ we have that $\lambda^2 = 1$. Note that both $\pm 1$ occur since 
we have $\Gamma_i (\Gamma_j \ket{\eta}) = - \lambda \Gamma_j \ket{\eta}$.
We can therefore express each $\Gamma_i$ as
$$
\Gamma_i = \Gamma_i^0 - \Gamma_i^1,
$$
where $\Gamma_i^0$ and $\Gamma_i^1$ are projectors onto the positive and
negative eigenspace of $\Gamma_i$ respectively. Furthermore, note that
we have for $i\neq j$
$$
\Tr(\Gamma_i \Gamma_j) = \frac{1}{2} \Tr(\Gamma_i \Gamma_j + \Gamma_j \Gamma_i) = 0.
$$
That is, all such operators are orthogonal.
To gain some intuition of why such operators may give good uncertainty relations note that
the positive and negative eigenspaces of such operators
are mutually unbiased (analogous to bases), since
for all $i \neq j$, and an arbitrary eigenvector $\ket{\psi_i}$
of $\Gamma_i$, 
$$
  \bra{\psi_i} \Gamma_j \ket{\psi_i} = 0\ .
$$
Hence, if we were measure the maximally mixed state on the positive eigenspace of $\Gamma_j$
with any of the other observables, the probability of obtaining a measurement outcome of 0 is
the same as for obtaining outcome 1.
For simplicity, we will write $H_\alpha(\Gamma_j|\rho) \assign H_\alpha(\{\Gamma_j^0,\Gamma_j^1\}|\rho)$.

It was shown~\cite{sa:clifford} that the following maximally strong uncertainty relation holds 
for any set of anti-commuting observables $\setS \subseteq \{\Gamma_j \mid j \in \{0,\ldots,2n\}\}$
$$
\min_{\rho} \frac{1}{|\setS|} \sum_{\Gamma_j \in \setS} H(\Gamma_j|\rho) = 1 - \frac{1}{|\setS|}\ .
$$
For dimension $d=2$, this reduces to an uncertainty relation for the mutually unbiased bases given by the eigenvectors
of $X$, $Z$ and $Y$ respectively.
For the collision entropy, the bound becomes
$$
\min_{\rho} \frac{1}{|\setS|} \sum_{\Gamma_j \in \setS} H_2\left(\Gamma_j|\rho\right)
      = 1 - \log\left( 1+\frac{1}{|\setS|} \right)
      \sim 1 - \frac{\log e}{|\setS|}\ ,
$$
and for the min-entropy we have
\begin{align}\label{eq:minEntropy}
\min_{\rho} \frac{1}{|\setS|} \sum_{\Gamma_j \in \setS} H_\infty\left(\Gamma_j|\rho\right)
      = 1 - \log\left( 1+\frac{1}{\sqrt{|\setS|}} \right)\ .
\end{align}
Interestingly, uncertainty relations for anti-commuting observables can also be
used to prove Tsirelson's bound~\cite{greg:relaxedUR}.

It is not known how to extend this result to more than two measurement outcomes. One may conjecture that the 
generalized Clifford algebra generated by 
operators $\Lambda_1,\ldots,\Lambda_n$, where for all $i \neq j$ we have 
$$
\Lambda_i \Lambda_j = \omega \Lambda_j \Lambda_i,
$$
with $\omega = e^{2 \Pi i/\ell}$ may give strong uncertainty relations for measurements with $\ell$ measurement outcomes. However, 
the example for $X_3$, $Z_3$ and $X_3 Z_3$ given above, and numerical evidence for higher dimensions refute this conjecture.

\section{Applications}\label{sec:apps}

Uncertainty relations for measurements in different bases have recently played an important role in proving security
of cryptographic protocols in the bounded~\cite{serge:new} and noisy-storage model~\cite{prl:noisy,noisy:new} respectively.
Here, uncertainty relations are used to bound the information that a cheating party has about bits which are encoded
into several possible bases, where the choice of basis is initially unknown to him. The simplest example is an encoding of a 
single bit $x_j \in \01$ into either the computational (as $\ket{x_i}$) or Hadamard basis (as $H\ket{x_j}$). Suppose
we choose the bit $x_j$, as well as the basis uniformly at random, and suppose further that 
the cheating party is allowed to perform any measurement on the encoded qubit giving him some classical information $K$.
After his measurement, we provide him with the basis information $\Theta$.
It can be shown using a purification argument, that we can turn the 
uncertainty relation for the min-entropy for the computational $\mB_1$ and Hadamard basis $\mB_2$ (see~\eqref{eq:minEntropy})
$$
\frac{1}{2}\left(H_\infty(\mB_1|\rho) + H_\infty(\mB_2|\rho)\right) \geq - \log \left(\frac{1}{2} + \frac{1}{2\sqrt{2}}\right)\ ,
$$
into the following bound for the adversary's knowledge about the bit $X_j$ given $K$ and the basis information $\Theta$
$$
H_\infty(X_j|K \Theta) \geq - \log\left(\frac{1}{2} + \frac{1}{2\sqrt{2}}\right)\ .
$$
The conditional min-entropy thereby has a very intuitive 
interpretation as $H_\infty(X_j|K \Theta) = - \log P_{\rm guess}(X_j| K \Theta)$,
where $P_{\rm guess}(X_j|K \Theta)$ is the average probability that the cheating party can guess $X_j$ given $K$ and $\Theta$, maximized over all 
strategies~\cite{robert:interpret}.

In a cryptographic setting, we are especially interested in the case where we repeat the encoding above many times. Suppose
we choose an $n$-bit string $X_1,\ldots,X_n$ uniformly at random, and encode each bit in either the computational or Hadamard
basis, also chosen uniformly and independently at random. Using the SDP formalism of~\cite{ww:pistar} it is easily seen~\cite{prl:noisy} that this gives
us
$$
H_\infty(X_1,\ldots,X_n|K, \Theta) \geq - n \log\left(\frac{1}{2} + \frac{1}{2\sqrt{2}}\right)\ .
$$
In the limit of large $n$, it is known that for independent states, the min-entropy behaves approximately like the Shannon
entropy~\cite{renato:diss,tomamichel:aep}. This allows one to turn the uncertainty relation of~\cite{maassen:ur} for the Shannon entropy 
into a better bound on the adversaries knowledge about the long string $X_1,\ldots,X_n$ in terms of the min-entropy. 
More precisely, it is 
known~\cite{serge:new} that
$$
H_\infty^\epsilon(X_1,\ldots,X_n|K, \Theta) \geq \left(\frac{1}{2} - 2 \delta\right)n
$$
for $\epsilon = \exp(- \delta^2 n /(32 (2 + \log(1/\delta))^2))$, where 
$H_\infty^\epsilon$ is the $\epsilon$-smooth min-entropy defined in~\cite{renato:diss}.
Intuitively, this quantity behaves like the min-entropy, except with
probability $\epsilon$. We refer to~\cite{noisy:new} for more information,
where this uncertainty relation was recently used to prove
security in the noisy-storage model.

\section{Open problems}

Since a full set of mutually unbiased bases form a $2$-design, it may be interesting to consider sets of
bases forming a $t$-design for any $t > 2$. Using the result of~\cite{klappenecker:designs} and the technique of~\cite{BallesterWehner}
it is straightforward to prove an incredibly weak uncertainty relation for the R{\'e}nyi entropy of order $t$, where the lower bound
obeys $1/(1-t) \log ((t! d!)/(t+d-1)!)$. Evidently, this lower bound becomes weaker for higher values of $t$, which is exactly the opposite of 
what one would hope for. It is an interesting open question, whether one can find good uncertainty relations for higher designs.

The most interesting open problem, however, is to find any sets of measurements at all
for which we do obtain maximally strong uncertainty relations for more than two measurement settings, and a constant
 number of measurement outcomes $|\setX| > 2$.
Note that always
\begin{equation}
  \label{eq:h-bounds}
  0 \leq c_{\{\setM_j\}} \leq \left( 1-\frac{1}{L} \right) \log |\setX|,
\end{equation}
for any set of measurements $\{\setM_j\}$ with outcomes in the set $\setX$.
The problem of the entropic uncertainty relations at its most
general is to find an expression, or at least a lower bound, for the quantity $c_{\{\setM_j\}}$
in ``simple'' terms of the geometry of the measurements $\setM_j$.

For measurements in different bases, which are of special interest for example in locking
applications~\cite{barbara:locking}, one is interested in the quantity
\[
  h(d;L) \assign \max_{\mB_1,\ldots,\mB_L} \min_{\rho} \frac{1}{L} \sum_{j=1}^L H(\mB_j|\rho)\ ,
\]
where the maximization is taken over bases $\mB_1,\ldots,\mB_L$.
Note that if in dimension $d$ there exist
$L$ mutually unbiased bases, then by virtue of~\eqref{eq:mubTrivial}
and the above (\ref{eq:h-bounds}),
\[
  \frac{1}{2}\log d \leq h(d;L) \leq \left( 1-\frac{1}{L} \right)\log d,
\]
and one would like to have a characterization of the sets of bases
attaining the maximum.

Seeing thus the scaling of $h(d;L)$ with $\log d$, and assuming an asymptotic viewpoint
of large dimension, we finally consider the 
quantity\footnote{If the limit exists; otherwise take the $\liminf$ or
$\limsup$, giving rise to $\underline{h}(L)$ and $\overline{h}(L)$,
respectively.} 
$$
  h(L) := \lim_{d\rightarrow\infty} \frac{1}{\log d} h(d;L)\ ,
$$
which depends now
only on the number of bases $L$. For example, $h(2) = 1/2$, and it is clear that
\[
  h(L+L') \geq \frac{L}{L+L'} h(L) + \frac{L'}{L+L'} h(L'),
\]
but we don't know if $h(L)$ actually strictly grows with $L$. If so, does
it approach the value $1-1/L$ suggested by the upper bound, or at least
$1-1/f(L)$ with some growing function $f$ of $L$?

\bibliographystyle{apsrmp}

\begin{thebibliography}{62}
\expandafter\ifx\csname natexlab\endcsname\relax\def\natexlab#1{#1}\fi
\expandafter\ifx\csname bibnamefont\endcsname\relax
  \def\bibnamefont#1{#1}\fi
\expandafter\ifx\csname bibfnamefont\endcsname\relax
  \def\bibfnamefont#1{#1}\fi
\expandafter\ifx\csname citenamefont\endcsname\relax
  \def\citenamefont#1{#1}\fi
\expandafter\ifx\csname url\endcsname\relax
  \def\url#1{\texttt{#1}}\fi
\expandafter\ifx\csname urlprefix\endcsname\relax\def\urlprefix{URL }\fi
\providecommand{\bibinfo}[2]{#2}
\providecommand{\eprint}[2][]{\url{#2}}

\bibitem[{\citenamefont{Ambainis}(2006)}]{andris}
\bibinfo{author}{\bibnamefont{Ambainis}, \bibfnamefont{A.}},
  \bibinfo{year}{2006}, \bibinfo{note}{in preparation.}

\bibitem[{\citenamefont{Azarchs}(2004)}]{azarchs:entropy}
\bibinfo{author}{\bibnamefont{Azarchs}, \bibfnamefont{A.}},
  \bibinfo{year}{2004}, \bibinfo{title}{Entropic uncertainty relations for
  incomplete sets of mutually unbiased observables},
  \bibinfo{note}{quant-ph/0412083}.

\bibitem[{\citenamefont{Ballester and Wehner}(2007)}]{BallesterWehner}
\bibinfo{author}{\bibnamefont{Ballester}, \bibfnamefont{M.}}, and
  \bibinfo{author}{\bibfnamefont{S.}~\bibnamefont{Wehner}},
  \bibinfo{year}{2007}, \bibinfo{journal}{Physical Review A}
  \textbf{\bibinfo{volume}{75}}, \bibinfo{pages}{022319}.

\bibitem[{\citenamefont{Ballester} \emph{et~al.}(2008)\citenamefont{Ballester,
  Wehner, and Winter}}]{ww:pistar}
\bibinfo{author}{\bibnamefont{Ballester}, \bibfnamefont{M.}},
  \bibinfo{author}{\bibfnamefont{S.}~\bibnamefont{Wehner}}, and
  \bibinfo{author}{\bibfnamefont{A.}~\bibnamefont{Winter}},
  \bibinfo{year}{2008}, \bibinfo{journal}{{IEEE} Transactions on Information
  Theory} \textbf{\bibinfo{volume}{54}}(\bibinfo{number}{9}),
  \bibinfo{pages}{4183}.

\bibitem[{\citenamefont{Bandyopadhyay}
  \emph{et~al.}(2002)\citenamefont{Bandyopadhyay, Boykin, Roychowdhury, and
  Vatan}}]{boykin:mub}
\bibinfo{author}{\bibnamefont{Bandyopadhyay}, \bibfnamefont{S.}},
  \bibinfo{author}{\bibfnamefont{P.}~\bibnamefont{Boykin}},
  \bibinfo{author}{\bibfnamefont{V.}~\bibnamefont{Roychowdhury}}, and
  \bibinfo{author}{\bibfnamefont{F.}~\bibnamefont{Vatan}},
  \bibinfo{year}{2002}, \bibinfo{journal}{Algorithmica}
  \textbf{\bibinfo{volume}{34}}(\bibinfo{number}{4}), \bibinfo{pages}{512}.

\bibitem[{\citenamefont{Beckner}(1975)}]{becker:ur}
\bibinfo{author}{\bibnamefont{Beckner}, \bibfnamefont{W.}},
  \bibinfo{year}{1975}, \bibinfo{journal}{Annals of Mathematics}
  \textbf{\bibinfo{volume}{102}}(\bibinfo{number}{1}), \bibinfo{pages}{159}.

\bibitem[{\citenamefont{Bia\l{}ynicki-Birula}(1984)}]{birula:angleUR}
\bibinfo{author}{\bibnamefont{Bia\l{}ynicki-Birula}, \bibfnamefont{I.}},
  \bibinfo{year}{1984}, \bibinfo{journal}{Physics Letters A}
  \textbf{\bibinfo{volume}{103}}(\bibinfo{number}{5}), \bibinfo{pages}{253}.

\bibitem[{\citenamefont{Bia\l{}ynicki-Birula}(2006)}]{birula:new}
\bibinfo{author}{\bibnamefont{Bia\l{}ynicki-Birula}, \bibfnamefont{I.}},
  \bibinfo{year}{2006}, \bibinfo{journal}{Physical Review A}
  \textbf{\bibinfo{volume}{74}}, \bibinfo{pages}{052102}.

\bibitem[{\citenamefont{Bia\l{}ynicki-Birula and
  Madajczyk}(1985)}]{birula:angularNew}
\bibinfo{author}{\bibnamefont{Bia\l{}ynicki-Birula}, \bibfnamefont{I.}}, and
  \bibinfo{author}{\bibfnamefont{J.~L.} \bibnamefont{Madajczyk}},
  \bibinfo{year}{1985}, \bibinfo{journal}{Physics Letters A}
  \textbf{\bibinfo{volume}{108}}(\bibinfo{number}{8}), \bibinfo{pages}{384}.

\bibitem[{\citenamefont{Bia\l{}ynicki-Birula and Mycielski}(1975)}]{BB:M}
\bibinfo{author}{\bibnamefont{Bia\l{}ynicki-Birula}, \bibfnamefont{I.}}, and
  \bibinfo{author}{\bibfnamefont{J.}~\bibnamefont{Mycielski}},
  \bibinfo{year}{1975}, \bibinfo{journal}{Communications in Mathematical
  Physics} \textbf{\bibinfo{volume}{44}}(\bibinfo{number}{129}).

\bibitem[{\citenamefont{Christandl and Winter}(2005)}]{CW05}
\bibinfo{author}{\bibnamefont{Christandl}, \bibfnamefont{M.}}, and
  \bibinfo{author}{\bibfnamefont{A.}~\bibnamefont{Winter}},
  \bibinfo{year}{2005}, \bibinfo{title}{Uncertainty, monogamy and locking of
  quantum correlations}, \bibinfo{note}{quant-ph/0501090}.

\bibitem[{\citenamefont{Damgaard} \emph{et~al.}(2007)\citenamefont{Damgaard,
  Fehr, Renner, Salvail, and Schaffner}}]{serge:new}
\bibinfo{author}{\bibnamefont{Damgaard}, \bibfnamefont{I.}},
  \bibinfo{author}{\bibfnamefont{S.}~\bibnamefont{Fehr}},
  \bibinfo{author}{\bibfnamefont{R.}~\bibnamefont{Renner}},
  \bibinfo{author}{\bibfnamefont{L.}~\bibnamefont{Salvail}}, and
  \bibinfo{author}{\bibfnamefont{C.}~\bibnamefont{Schaffner}},
  \bibinfo{year}{2007}, \bibinfo{title}{A tight high-order entropic uncertainty
  relation with applications in the bounded quantum-storage model},
  \bibinfo{note}{proceedings of CRYPTO 2007}.

\bibitem[{\citenamefont{Damgaard} \emph{et~al.}(2005)\citenamefont{Damgaard,
  Fehr, Salvail, and Schaffner}}]{serge:bounded}
\bibinfo{author}{\bibnamefont{Damgaard}, \bibfnamefont{I.}},
  \bibinfo{author}{\bibfnamefont{S.}~\bibnamefont{Fehr}},
  \bibinfo{author}{\bibfnamefont{L.}~\bibnamefont{Salvail}}, and
  \bibinfo{author}{\bibfnamefont{C.}~\bibnamefont{Schaffner}},
  \bibinfo{year}{2005}, in \emph{\bibinfo{booktitle}{Proceedings of 46th IEEE
  FOCS}}, pp. \bibinfo{pages}{449--458}.

\bibitem[{\citenamefont{Deutsch}(1983)}]{deutsch:ur}
\bibinfo{author}{\bibnamefont{Deutsch}, \bibfnamefont{D.}},
  \bibinfo{year}{1983}, \bibinfo{journal}{Physical Review Letters}
  \textbf{\bibinfo{volume}{50}}, \bibinfo{pages}{631}.

\bibitem[{\citenamefont{Dietz}(2006)}]{dietz:blochsphere}
\bibinfo{author}{\bibnamefont{Dietz}, \bibfnamefont{K.}}, \bibinfo{year}{2006},
  \bibinfo{journal}{Journal of Physics A: Math. Gen.}
  \textbf{\bibinfo{volume}{36}}(\bibinfo{number}{6}), \bibinfo{pages}{1433}.

\bibitem[{\citenamefont{DiVincenzo}
  \emph{et~al.}(2004)\citenamefont{DiVincenzo, Horodecki, Leung, Smolin, and
  Terhal}}]{barbara:locking}
\bibinfo{author}{\bibnamefont{DiVincenzo}, \bibfnamefont{D.}},
  \bibinfo{author}{\bibfnamefont{M.}~\bibnamefont{Horodecki}},
  \bibinfo{author}{\bibfnamefont{D.}~\bibnamefont{Leung}},
  \bibinfo{author}{\bibfnamefont{J.}~\bibnamefont{Smolin}}, and
  \bibinfo{author}{\bibfnamefont{B.}~\bibnamefont{Terhal}},
  \bibinfo{year}{2004}, \bibinfo{journal}{Physical Review Letters}
  \textbf{\bibinfo{volume}{92}}(\bibinfo{number}{067902}).

\bibitem[{\citenamefont{Ghirardi} \emph{et~al.}(2003)\citenamefont{Ghirardi,
  Marinatto, and Romano}}]{ghirardi:ur}
\bibinfo{author}{\bibnamefont{Ghirardi}, \bibfnamefont{G.}},
  \bibinfo{author}{\bibfnamefont{L.}~\bibnamefont{Marinatto}}, and
  \bibinfo{author}{\bibfnamefont{R.}~\bibnamefont{Romano}},
  \bibinfo{year}{2003}, \bibinfo{journal}{Physics Letters A}
  \textbf{\bibinfo{volume}{317}}(\bibinfo{number}{1}), \bibinfo{pages}{32}.

\bibitem[{\citenamefont{Grassl}(2004)}]{grassl:mub}
\bibinfo{author}{\bibnamefont{Grassl}, \bibfnamefont{M.}},
  \bibinfo{year}{2004}, in \emph{\bibinfo{booktitle}{Proceedings ERATO
  Conference on Quantum Information Science}}, pp. \bibinfo{pages}{60--61},
  \eprint{quant-ph/0406175}.

\bibitem[{\citenamefont{Guehne}(2004)}]{guehne:separable}
\bibinfo{author}{\bibnamefont{Guehne}, \bibfnamefont{O.}},
  \bibinfo{year}{2004}, \bibinfo{journal}{Physical Review Letters}
  \textbf{\bibinfo{volume}{92}}, \bibinfo{pages}{117903}.

\bibitem[{\citenamefont{Hall}(1997)}]{hall:ur}
\bibinfo{author}{\bibnamefont{Hall}, \bibfnamefont{M.~J.~W.}},
  \bibinfo{year}{1997}, \bibinfo{journal}{Physical Review A}
  \textbf{\bibinfo{volume}{55}}, \bibinfo{pages}{100}.

\bibitem[{\citenamefont{Hayden} \emph{et~al.}(2004)\citenamefont{Hayden, Leung,
  Shor, and Winter}}]{rand}
\bibinfo{author}{\bibnamefont{Hayden}, \bibfnamefont{P.}},
  \bibinfo{author}{\bibfnamefont{D.}~\bibnamefont{Leung}},
  \bibinfo{author}{\bibfnamefont{P.}~\bibnamefont{Shor}}, and
  \bibinfo{author}{\bibfnamefont{A.}~\bibnamefont{Winter}},
  \bibinfo{year}{2004}, \bibinfo{journal}{Communications in Mathematical
  Physics} \textbf{\bibinfo{volume}{250}}(\bibinfo{number}{2}),
  \bibinfo{pages}{371}.

\bibitem[{\citenamefont{Heisenberg}(1927)}]{heisenberg:ur}
\bibinfo{author}{\bibnamefont{Heisenberg}, \bibfnamefont{W.}},
  \bibinfo{year}{1927}, \bibinfo{journal}{Zeitschrift f{\"u}r Physik}
  \textbf{\bibinfo{volume}{43}}, \bibinfo{pages}{172}.

\bibitem[{\citenamefont{Hirschmann}(1957)}]{hirschmann:ur}
\bibinfo{author}{\bibnamefont{Hirschmann}, \bibfnamefont{I.~I.}},
  \bibinfo{year}{1957}, \bibinfo{journal}{American Journal of Mathematics}
  \textbf{\bibinfo{volume}{79}}(\bibinfo{number}{152}).

\bibitem[{\citenamefont{Ivanovic}(1992)}]{ivanovic:ur}
\bibinfo{author}{\bibnamefont{Ivanovic}, \bibfnamefont{I.~D.}},
  \bibinfo{year}{1992}, \bibinfo{journal}{J. Phys. A: Math. Gen.}
  \textbf{\bibinfo{volume}{25}}(\bibinfo{number}{7}), \bibinfo{pages}{363}.

\bibitem[{\citenamefont{Jordan and Wigner}(1928)}]{JordanWigner}
\bibinfo{author}{\bibnamefont{Jordan}, \bibfnamefont{P.}}, and
  \bibinfo{author}{\bibfnamefont{E.}~\bibnamefont{Wigner}},
  \bibinfo{year}{1928}, \bibinfo{journal}{Zeitschrift f{\"u}r Physik}
  \textbf{\bibinfo{volume}{47}}, \bibinfo{pages}{631}.

\bibitem[{\citenamefont{Klappenecker and
  R{\"o}tteler}(2004)}]{klappenecker:mubs}
\bibinfo{author}{\bibnamefont{Klappenecker}, \bibfnamefont{A.}}, and
  \bibinfo{author}{\bibfnamefont{M.}~\bibnamefont{R{\"o}tteler}},
  \bibinfo{year}{2004}, in \emph{\bibinfo{booktitle}{International Conference
  on Finite Fields and Applications (Fq7)}} (\bibinfo{publisher}{Springer}),
  volume \bibinfo{volume}{2948} of \emph{\bibinfo{series}{Lecture Notes in
  Computer Science}}, pp. \bibinfo{pages}{137--144}.

\bibitem[{\citenamefont{Klappenecker and
  R{\"{o}}tteler}(2005)}]{klappenecker:designs}
\bibinfo{author}{\bibnamefont{Klappenecker}, \bibfnamefont{A.}}, and
  \bibinfo{author}{\bibfnamefont{M.}~\bibnamefont{R{\"{o}}tteler}},
  \bibinfo{year}{2005}, in \emph{\bibinfo{booktitle}{Proceedings of IEEE
  International Symposium on Information Theory}}, pp.
  \bibinfo{pages}{1740--1744}.

\bibitem[{\citenamefont{Koashi}(2005)}]{qkd:ur}
\bibinfo{author}{\bibnamefont{Koashi}, \bibfnamefont{M.}},
  \bibinfo{year}{2005}, \bibinfo{title}{Simple security proof of quantum key
  distribution via uncertainty principle}, \bibinfo{note}{quant-ph/0505108}.

\bibitem[{\citenamefont{K\"onig} \emph{et~al.}(2008)\citenamefont{K\"onig,
  Renner, and Schaffner}}]{robert:interpret}
\bibinfo{author}{\bibnamefont{K\"onig}, \bibfnamefont{R.}},
  \bibinfo{author}{\bibfnamefont{R.}~\bibnamefont{Renner}}, and
  \bibinfo{author}{\bibfnamefont{C.}~\bibnamefont{Schaffner}},
  \bibinfo{year}{2008}, \bibinfo{title}{The operational meaning of min- and
  max-entropy}, \bibinfo{note}{arXiv:0807.1338}.

\bibitem[{\citenamefont{K{\"o}nig} \emph{et~al.}(2009)\citenamefont{K{\"o}nig,
  Wehner, and Wullschleger}}]{noisy:new}
\bibinfo{author}{\bibnamefont{K{\"o}nig}, \bibfnamefont{R.}},
  \bibinfo{author}{\bibfnamefont{S.}~\bibnamefont{Wehner}}, and
  \bibinfo{author}{\bibfnamefont{J.}~\bibnamefont{Wullschleger}},
  \bibinfo{year}{2009}, \bibinfo{title}{Unconditional security in the
  noisy-storage model}, \bibinfo{note}{arXiv:0906.1030}.

\bibitem[{\citenamefont{Kraus}(1987)}]{kraus:ur}
\bibinfo{author}{\bibnamefont{Kraus}, \bibfnamefont{K.}}, \bibinfo{year}{1987},
  \bibinfo{journal}{Physical Review D}
  \textbf{\bibinfo{volume}{35}}(\bibinfo{number}{10}), \bibinfo{pages}{3070}.

\bibitem[{\citenamefont{Krishna and Parthasarathy}(2002)}]{krishna:ur}
\bibinfo{author}{\bibnamefont{Krishna}, \bibfnamefont{M.}}, and
  \bibinfo{author}{\bibfnamefont{K.~R.} \bibnamefont{Parthasarathy}},
  \bibinfo{year}{2002}, \bibinfo{journal}{Indian J. of Statistics Ser. A}
  \textbf{\bibinfo{volume}{64}}(\bibinfo{number}{842}),
  \bibinfo{note}{quant-ph/0110025}.

\bibitem[{\citenamefont{Landau and Pollack}(1961)}]{landau:ur}
\bibinfo{author}{\bibnamefont{Landau}, \bibfnamefont{H.~J.}}, and
  \bibinfo{author}{\bibfnamefont{H.~O.} \bibnamefont{Pollack}},
  \bibinfo{year}{1961}, \bibinfo{journal}{Bell Syst. Tech. J.}
  \textbf{\bibinfo{volume}{40}}(\bibinfo{number}{65}).

\bibitem[{\citenamefont{Larsen}(1990)}]{larsen:entropy}
\bibinfo{author}{\bibnamefont{Larsen}, \bibfnamefont{U.}},
  \bibinfo{year}{1990}, \bibinfo{journal}{J. Phys. A: Math. Gen.}
  \textbf{\bibinfo{volume}{23}}, \bibinfo{pages}{1041}.

\bibitem[{\citenamefont{Lawrence} \emph{et~al.}(2002)\citenamefont{Lawrence,
  Brukner, and Zeilinger}}]{lawrence:mub}
\bibinfo{author}{\bibnamefont{Lawrence}, \bibfnamefont{J.}},
  \bibinfo{author}{\bibfnamefont{C.}~\bibnamefont{Brukner}}, and
  \bibinfo{author}{\bibfnamefont{A.}~\bibnamefont{Zeilinger}},
  \bibinfo{year}{2002}, \bibinfo{journal}{Physical Review A}
  \textbf{\bibinfo{volume}{65}}, \bibinfo{pages}{032320}.

\bibitem[{\citenamefont{Lounesto}(2001)}]{lounesto:book}
\bibinfo{author}{\bibnamefont{Lounesto}, \bibfnamefont{P.}},
  \bibinfo{year}{2001}, \emph{\bibinfo{title}{Clifford Algebras and Spinors}}
  (\bibinfo{publisher}{Cambridge University Press}).

\bibitem[{\citenamefont{Maassen and Uffink}(1988)}]{maassen:ur}
\bibinfo{author}{\bibnamefont{Maassen}, \bibfnamefont{H.}}, and
  \bibinfo{author}{\bibfnamefont{J.}~\bibnamefont{Uffink}},
  \bibinfo{year}{1988}, \bibinfo{journal}{Physical Review Letters}
  \textbf{\bibinfo{volume}{60}}(\bibinfo{number}{1103}).

\bibitem[{\citenamefont{Massar}(2007)}]{massar:ur}
\bibinfo{author}{\bibnamefont{Massar}, \bibfnamefont{S.}},
  \bibinfo{year}{2007}, \bibinfo{journal}{Physical Review A}
  \textbf{\bibinfo{volume}{76}}, \bibinfo{pages}{042114}.

\bibitem[{\citenamefont{Partovi}(1983)}]{partovi:angleUR}
\bibinfo{author}{\bibnamefont{Partovi}, \bibfnamefont{M.~H.}},
  \bibinfo{year}{1983}, \bibinfo{journal}{Physical Review Letters}
  \textbf{\bibinfo{volume}{50}}(\bibinfo{number}{24}).

\bibitem[{\citenamefont{Rastegin}(2008{\natexlab{a}})}]{rastegin:errMassar}
\bibinfo{author}{\bibnamefont{Rastegin}, \bibfnamefont{A.~E.}},
  \bibinfo{year}{2008}{\natexlab{a}}, \bibinfo{title}{Comment on ``uncertainty
  relations for positive-operator-valued measures``},
  \bibinfo{note}{arxiv:0810.0038}.

\bibitem[{\citenamefont{Rastegin}(2008{\natexlab{b}})}]{rastegin:ur2}
\bibinfo{author}{\bibnamefont{Rastegin}, \bibfnamefont{A.~E.}},
  \bibinfo{year}{2008}{\natexlab{b}}, \bibinfo{note}{arXiv:0807.2691}.

\bibitem[{\citenamefont{Rastegin}(2008{\natexlab{c}})}]{rastegin:ur1}
\bibinfo{author}{\bibnamefont{Rastegin}, \bibfnamefont{A.~E.}},
  \bibinfo{year}{2008}{\natexlab{c}}, \bibinfo{note}{arXiv:0805.1777}.

\bibitem[{\citenamefont{Renes and Boileau}(2009)}]{RenesBoileau:inequality}
\bibinfo{author}{\bibnamefont{Renes}, \bibfnamefont{J.~M.}}, and
  \bibinfo{author}{\bibfnamefont{J.-C.} \bibnamefont{Boileau}},
  \bibinfo{year}{2009}, \bibinfo{journal}{Physical Review Letters}
  \textbf{\bibinfo{volume}{103}}, \bibinfo{pages}{020402}.

\bibitem[{\citenamefont{Renner}(2005)}]{renato:diss}
\bibinfo{author}{\bibnamefont{Renner}, \bibfnamefont{R.}},
  \bibinfo{year}{2005}, \emph{\bibinfo{title}{Security of Quantum Key
  Distribution}}, Ph.D. thesis, \bibinfo{school}{ETH Zurich},
  \bibinfo{note}{quant-ph/0512258}.

\bibitem[{\citenamefont{R{\'e}nyi}(1960)}]{renyi:entropy}
\bibinfo{author}{\bibnamefont{R{\'e}nyi}, \bibfnamefont{A.}},
  \bibinfo{year}{1960}, in \emph{\bibinfo{booktitle}{Proceedings of the 4th
  Berkeley Symposium on Mathematics, Statistics and Probability}}, pp.
  \bibinfo{pages}{547--561}.

\bibitem[{\citenamefont{Riesz}(1929)}]{riesz:norms}
\bibinfo{author}{\bibnamefont{Riesz}, \bibfnamefont{M.}}, \bibinfo{year}{1929},
  \bibinfo{journal}{Acta Math.}
  \textbf{\bibinfo{volume}{49}}(\bibinfo{number}{465}).

\bibitem[{\citenamefont{Robertson}(1929)}]{robinson:uncertainty}
\bibinfo{author}{\bibnamefont{Robertson}, \bibfnamefont{H.}},
  \bibinfo{year}{1929}, \bibinfo{journal}{Physical Review}
  \textbf{\bibinfo{volume}{34}}, \bibinfo{pages}{163}.

\bibitem[{\citenamefont{Sanchez}(1993)}]{sanchez:old}
\bibinfo{author}{\bibnamefont{Sanchez}, \bibfnamefont{J.}},
  \bibinfo{year}{1993}, \bibinfo{journal}{Physics Letters A}
  \textbf{\bibinfo{volume}{173}}, \bibinfo{pages}{233}.

\bibitem[{\citenamefont{Sanchez-Ruiz}(1995)}]{sanchez:improved}
\bibinfo{author}{\bibnamefont{Sanchez-Ruiz}, \bibfnamefont{J.}},
  \bibinfo{year}{1995}, \bibinfo{journal}{Physics Letters A}
  \textbf{\bibinfo{volume}{201}}, \bibinfo{pages}{125}.

\bibitem[{\citenamefont{Sanchez-Ruiz}(1998)}]{sanchez:entropyD2}
\bibinfo{author}{\bibnamefont{Sanchez-Ruiz}, \bibfnamefont{J.}},
  \bibinfo{year}{1998}, \bibinfo{journal}{Physics Letters A}
  \textbf{\bibinfo{volume}{244}}, \bibinfo{pages}{189}.

\bibitem[{\citenamefont{Shannon}(1948)}]{shannon:entropy}
\bibinfo{author}{\bibnamefont{Shannon}, \bibfnamefont{C.~E.}},
  \bibinfo{year}{1948}, \bibinfo{journal}{Bell System Technical Journal}
  \textbf{\bibinfo{volume}{27}}, \bibinfo{pages}{379}.

\bibitem[{\citenamefont{Tomamichel}
  \emph{et~al.}(2008)\citenamefont{Tomamichel, Colbeck, and
  Renner}}]{tomamichel:aep}
\bibinfo{author}{\bibnamefont{Tomamichel}, \bibfnamefont{M.}},
  \bibinfo{author}{\bibfnamefont{R.}~\bibnamefont{Colbeck}}, and
  \bibinfo{author}{\bibfnamefont{R.}~\bibnamefont{Renner}},
  \bibinfo{year}{2008}, \bibinfo{title}{A fully quantum asymptotic
  equipartition property}, \bibinfo{note}{arXiv:0811.1221}.

\bibitem[{\citenamefont{{Ver Steeg} and Wehner}(2009)}]{greg:relaxedUR}
\bibinfo{author}{\bibnamefont{{Ver Steeg}}, \bibfnamefont{G.}}, and
  \bibinfo{author}{\bibfnamefont{S.}~\bibnamefont{Wehner}},
  \bibinfo{year}{2009}, \bibinfo{journal}{Quantum Information and Computation}
  \textbf{\bibinfo{volume}{9}}, \bibinfo{pages}{801}.

\bibitem[{\citenamefont{de~Vicente and
  Sanchez-Ruiz}(2008)}]{sanchez:evenBetter}
\bibinfo{author}{\bibnamefont{de~Vicente}, \bibfnamefont{J.~I.}}, and
  \bibinfo{author}{\bibfnamefont{J.}~\bibnamefont{Sanchez-Ruiz}},
  \bibinfo{year}{2008}, \bibinfo{journal}{Physical Review A}
  \textbf{\bibinfo{volume}{77}}, \bibinfo{pages}{042110}.

\bibitem[{\citenamefont{Wehner}(2008)}]{steph:diss}
\bibinfo{author}{\bibnamefont{Wehner}, \bibfnamefont{S.}},
  \bibinfo{year}{2008}, \emph{\bibinfo{title}{Cryptography in a quantum
  world}}, Ph.D. thesis, \bibinfo{school}{University of Amsterdam},
  \bibinfo{note}{arXiv:0806.3483}.

\bibitem[{\citenamefont{Wehner} \emph{et~al.}(2008)\citenamefont{Wehner,
  Schaffner, and Terhal}}]{prl:noisy}
\bibinfo{author}{\bibnamefont{Wehner}, \bibfnamefont{S.}},
  \bibinfo{author}{\bibfnamefont{C.}~\bibnamefont{Schaffner}}, and
  \bibinfo{author}{\bibfnamefont{B.~M.} \bibnamefont{Terhal}},
  \bibinfo{year}{2008}, \bibinfo{journal}{Physical Review Letters}
  \textbf{\bibinfo{volume}{100}}(\bibinfo{number}{22}), \bibinfo{eid}{220502}.

\bibitem[{\citenamefont{Wehner and Winter}(2008)}]{sa:clifford}
\bibinfo{author}{\bibnamefont{Wehner}, \bibfnamefont{S.}}, and
  \bibinfo{author}{\bibfnamefont{A.}~\bibnamefont{Winter}},
  \bibinfo{year}{2008}, \bibinfo{journal}{Journal of Mathematical Physics}
  \textbf{\bibinfo{volume}{49}}, \bibinfo{pages}{062105}.

\bibitem[{\citenamefont{Wocjan and Beth}(2005)}]{wocjan:mub}
\bibinfo{author}{\bibnamefont{Wocjan}, \bibfnamefont{P.}}, and
  \bibinfo{author}{\bibfnamefont{T.}~\bibnamefont{Beth}}, \bibinfo{year}{2005},
  \bibinfo{journal}{Quantum Information and Computation}
  \textbf{\bibinfo{volume}{5}}(\bibinfo{number}{2}), \bibinfo{pages}{93}.

\bibitem[{\citenamefont{Wootters and Fields}(1989)}]{wootters:mub}
\bibinfo{author}{\bibnamefont{Wootters}, \bibfnamefont{W.}}, and
  \bibinfo{author}{\bibfnamefont{B.}~\bibnamefont{Fields}},
  \bibinfo{year}{1989}, \bibinfo{journal}{Ann. Phys.}
  \textbf{\bibinfo{volume}{191}}(\bibinfo{number}{368}).

\bibitem[{\citenamefont{Wootters and Sussman}(2007)}]{wootters:mur}
\bibinfo{author}{\bibnamefont{Wootters}, \bibfnamefont{W.~K.}}, and
  \bibinfo{author}{\bibfnamefont{D.~M.} \bibnamefont{Sussman}},
  \bibinfo{year}{2007}, \bibinfo{note}{arXiv:0704.1277}.

\bibitem[{\citenamefont{Wu} \emph{et~al.}(2009)\citenamefont{Wu, Yu, and
  Molmer}}]{wu:boundAgain}
\bibinfo{author}{\bibnamefont{Wu}, \bibfnamefont{S.}},
  \bibinfo{author}{\bibfnamefont{S.}~\bibnamefont{Yu}}, and
  \bibinfo{author}{\bibfnamefont{K.}~\bibnamefont{Molmer}},
  \bibinfo{year}{2009}, \bibinfo{journal}{Physical Review A}
  \textbf{\bibinfo{volume}{79}}, \bibinfo{pages}{022104}.

\bibitem[{\citenamefont{Zauner}(1999)}]{zauner:diss}
\bibinfo{author}{\bibnamefont{Zauner}, \bibfnamefont{G.}},
  \bibinfo{year}{1999}, \emph{\bibinfo{title}{Quantendesigns - Grundz{\"u}ge
  einer nichtkommutativen Designtheorie}}, Ph.D. thesis,
  \bibinfo{school}{Universit{\"a}t Wien}.

\end{thebibliography}

\appendix

\section{A bound for mutually unbiased bases}

Here we provide an alternative proof of an entropic uncertainty relation
for a full set of mutually unbiased bases in dimension $d=2^n$. This has previously been proved
in~\cite{sanchez:old,ivanovic:ur}. We already provided an alternative proof using
the fact that the set of all mutually unbiased bases forms a 
2-design~\cite{BallesterWehner}.
The present a very simple alternative proof for dimension $d=2^n$ which has the advantage that it neither requires the 
introduction of 2-designs, nor the results of~\cite{larsen:entropy} that were used in
the previous proof by Sanchez-Ruiz~\cite{sanchez:old}. Instead, our proof~\cite{steph:diss} 
is elementary: After choosing a convenient parametrization of quantum states, 
the statement follows immediately from Fourier analysis. 

For the parametrization,
we first introduce a basis for the space of $2^n \times 2^n$ matrices 
with the help of mutually unbiased bases. Recall 
that in dimension $2^n$, we can find exactly $2^n+1$ MUBs.
We will use the short-hand notation $[k] := \{1,\ldots,k\}$, and
write $j \oplus j'$ to denote the bitwise xor of strings $j$ and $j'$.

\begin{lemma}
Consider the Hermitian matrices
$$
S^j_b = \sum_{x \in \01^n} (-1)^{j \cdot x} \outp{x_b}{x_b},
$$
for $b \in [d+1]$,
$j \in [d-1]$ and for all $x,x' \in \01^n$ and
$b\neq b' \in [d+1]$ we have $|\inp{x_b}{x'_{b'}}|^2 = 1/d$. Then 
the set $\{\id\} \cup \{S^j_b\mid b \in [d+1], j \in [d-1]\}$ forms
a basis for the space of $d \times d$ matrices, where for all $j$ and $b$, $S^j_b$ is traceless and $(S^j_b)^2 = \id$.
\end{lemma}
\begin{proof}
First, note that we have $(d+1)(d-1) + 1 = d^2$ matrices. We now show that
they are all orthogonal. Note that
$$
\Tr(S^j_b) = \sum_{x\in\01^n} (-1)^{j \cdot x} = 0,
$$
since $j \neq 0$, and hence $S^j_b$ is traceless. Hence $\Tr(\id S^j_b) = 0$. Furthermore,
\begin{equation}\label{orthogonalBasis}
\Tr(S^j_b S^{j'}_{b'}) = \sum_{x,x' \in \01^n} (-1)^{j \cdot x} (-1)^{j' \cdot x'} |\inp{x_b}{x'_{b'}}|^2.
\end{equation}
For $b\neq b'$, Eq.~(\ref{orthogonalBasis}) gives us $\Tr(S^j_b S^{j'}_{b'}) = (1/d) \left(\sum_x (-1)^{j \cdot x}\right)
\left(\sum_{x'} (-1)^{j' \cdot x'}\right) = 0$, since $j,j' \neq 0$. For $b=b'$, but $j \neq j'$, we get
$\Tr(S^j_b S^{j'}_{b'})= \sum_x (-1)^{(j \oplus j') \cdot x} = 0$ since $j \oplus j' \neq 0$.

Finally, $\left(S^j_b\right)^2 = \sum_{xx'} (-1)^{j \cdot x}(-1)^{j \cdot x'} \outp{x_b}{x_b}\outp{x'_b}{x'_b} = \id$.
\end{proof}

Since $\{\id,S^j_b\}$ form a basis for the $d \times d$ matrices, we can thus express the state $\rho$ of a $d$-dimensional
system as
$$
\rho = \frac{1}{d}\left(\id + \sum_{b \in [d+1]}\sum_{j \in [d-1]} s^j_b S^j_b\right),
$$
for some coefficients $s^j_b \in \Real$. It is now easy to see that
\begin{lemma}\label{MUBpurity}
Let $\rho$ be a pure state parametrized as above. Then 
$$
\sum_{b \in [d+1]}\sum_{j \in [d-1]} (s^j_b)^2 = d-1.
$$
\end{lemma}
\begin{proof}
If $\rho$ is a pure state, we have $\Tr(\rho^2) = 1$. Hence
\begin{eqnarray*}
\Tr(\rho^2) &=& \frac{1}{d^2}\left(\Tr(\id) + \sum_{b \in [d+1]}
\sum_{j \in [d-1]} (s^j_b)^2 \Tr(\id)\right)\\
&=&\frac{1}{d}\left(1 + \sum_b \sum_j (s^j_b)^2\right) = 1,
\end{eqnarray*}
from which the claim follows.
\end{proof}
Suppose now that we are given a set of $d+1$ MUBs $\mB_1,\ldots,\mB_{d+1}$
with $\mB_b = \{\ket{x_b}\mid x \in \01^n\}$. Then the following simple observation lies at the core of our proof:
\begin{lemma}\label{MUBexpansion}
Let $\ket{x_b}$ be the $x$-th basis vector of the $b$-th MUB. Then for any state $\rho$
$$
\Tr(\outp{x_b}{x_b} \rho) = \frac{1}{d}\left(1 + \sum_{j \in [d-1} (-1)^{j \cdot x} s^j_{b}\right).
$$
\end{lemma}
\begin{proof}
We have
$$
\Tr(\outp{x_b}{x_b}\rho) = \frac{1}{d}\left(\Tr(\outp{x_b}{x_b}) + \sum_{b',j} s^j_{b'} \Tr(S^j_{b'} \outp{x_b}{x_b})\right)
$$
Suppose $b \neq b'$. Then $\Tr(S^j_{b'} \outp{x_b}{x_b}) = (1/d) \sum_{x'} (-1)^{j \cdot x'} = 0$, since $j \neq 0$.
Suppose $b = b'$. Then $\Tr(S^j_{b'} \outp{x_b}{x_b}) = \sum_{x'} (-1)^{j \cdot x'} |\inp{x_b}{x'_b}|^2 = (-1)^{j \cdot x}$, 
from which the claim follows.
\end{proof}

We are now ready to prove an entropic uncertainty relation for $L$ mutually unbiased bases.
\begin{theorem}\label{MUBmany}
Let $\mS = \{\mB_1,\ldots,\mB_L\}$ be a set of mutually unbiased bases. Then
$$
\frac{1}{L}\sum_{b \in [L]} H_2(\mB_b,\ket{\Psi}) \geq - \log \frac{L + d - 1}{d L}.
$$
\end{theorem}
\begin{proof}
First, note that we can define functions $f_b(j) = s^j_b$ for $j \in [d-1]$ and $f_b(0) = 1$. 
Then $\hat{f}_b(x) = (1/\sqrt{d})(\sum_{j\in\{0,\ldots,d-1\}} (-1)^{j \cdot x} s^j_b)$ is the Fourier
transform of $f_b$ and $(1/\sqrt{d}) \hat{f}_b(x) = \Tr(\outp{x_b}{x_b})$ by Lemma~\ref{MUBexpansion}.
Thus
\begin{eqnarray*}
\frac{1}{L}\sum_{b \in [L]} H_2(\mB_b,\ket{\Psi}) &=& -\frac{1}{L}\sum_{b \in [L]} \log \sum_{x \in \01^n} 
|\inp{x_b}{\Psi}|^4\\
&\geq& - \log \frac{1}{d L} \sum_b \sum_x \hat{f}_b(x)^2\\
&=& - \log \frac{1}{d L} \sum_b \left(1 + \sum_j (s^j_b)^2\right)\\
&=& - \log \frac{1}{d L} (L + d - 1),
\end{eqnarray*}
where the first inequality follows from Jensen's inequality and the concavity of $\log$. The next equality
follows from Parseval's equality, and the last follows from the fact that $\ket{\Psi}$ is a pure state
and Lemma~\ref{MUBpurity}.
\end{proof}

\begin{corollary}
Let $\mS = \{\mB_1,\ldots,\mB_L\}$ be a set of mutually unbiased bases. Then
$$
\frac{1}{L}\sum_{b \in [L]} H(\mB_b|\ket{\Psi}) \geq - \log \frac{L + d - 1}{d L }.
$$
In particular, for a full set of $L = d+1$ MUBs we get
$\frac{1}{L} \sum_b H(\mB_b|\ket{\Psi}) \geq \log(d+1) - 1$.
\end{corollary}
\begin{proof}
This follows immediately from Theorem~\ref{MUBmany} and the fact that $H(\cdot) \geq H_2(\cdot)$.
\end{proof}

It is interesting to note that this bound is the same that arises from interpolating between 
the results of Sanchez-Ruiz~\cite{sanchez:old} and Maassen and Uffink~\cite{maassen:ur}
as was done by Azarchs~\cite{azarchs:entropy}. This bound has more recently been rediscovered by~\cite{wu:boundAgain}.

\end{document}